# Impact of Information Technology in Cyberwars: Systematic Literature Review

Santhosh, Pogaku

Student – MSIT School of Information Technology, University of Cincinnati, Pogakush @mail.uc.edu

## 1 ABSTRACT

Different types of warfare have evolved between nations and states in the modern era, each with its technological breakthroughs and use of cutting-edge technologies. With the help of the latest innovations, technologies and ideas emerging and contributing more to the It sector, making it more advanced and resulting in different technologies used for cyber warfare, information technology has a stronghold, power, and control over many other integrated automated technologies. To identify the various technologies that are primarily used in cyber warfare. This exploratory study used a systematic review technique and a theme analysis approach to examine prior works in information technology relevant to cyber warfare.

## 2 INTRODUCTION

In this study, we explore the societal phenomenon of how information technology is used in cyber warfare and how it plays a vital role using a systematic literature review method. Specific studies in the digital databases such as IEEE and ACM about IT playing a crucial role in the cyber ware fare. In recent times, cyber-attacks have been made on the defence system and organizations, which led to huge losses [1]. The Author observed the use of IT in Cyber-warfare after noticing the widespread use of information technology in combat. Cyber assaults or attacks are unwelcome attempts to steal, expose, alter, disable, or destroy information by gaining unauthorized access to computer systems. [2], and there is a specific factor of losses such as time, data, and resources if an organization has been cyber-attacked. The Author discussed a study by Thierry Mbah and Barry  Dwolatzky [3]  about the threat to the banking system in South Africa. The growth of cyber-crimes is becoming a serious economic issue for South African organizations and the country. This paper begins by analyzing the various cyber threats involved in cyberbanking. A study by Jabber Ahmed and Quddus [4] examines the cyber-attacks during a covid-19 pandemicFrom a cyber-crime aspect, specific cyber-attacks occurred globally during the COVID-19 pandemic, illustrating the broad breadth of cyber-attacks that occurred worldwide during the epidemic. A finding by A. Salih, S. T. Zeebaree, S. Ameen, A. Alkhyyat and H. M. Shukur [5] related to the growing technology in the markets the various tools such as Artificial intelligence and machine learning play a vital role to launch the cyberwar, and this article is very much related to our research study. The study by Puchong li and Qinghui Liu [6] examines and reviews the usual advances in cyber security and looks at the challenges, shortcomings, and benefits of the various approaches. The many types of new descendant attacks are thoroughly addressed.. The article by C. Iwendi et al. [7] protects against cyber-attacks, and one article published by zero watermarking algorithms. The study by Sylvain P. Leblanc [8] provides an overview of the current state of the art in cyber-attack simulation and modeling, as well as defensive countermeasures. It encompasses a wide range of cyber warfare simulations, including realistic, virtual, and constructive scenarios. As per James P and Rafal [9], the discovery in June 2010 that a computer worm is known as 'Stuxnet' had attacked Iran's nuclear plant at Natanz hinted that the future of cyber warfare was today. The political and strategic framework in which new cyber threats emerge and the consequences the worm has had in this regard are even more critical. The study by Adam and Andrew [10] on cyber security information sharing enumerates a long number of possible benefits for both public and private sector companies. Despite the apparent benefits, successful cyber security information exchange has proven difficult. Some articles explained the automated cyber threat sensing and acting towards it, and one is by Peter Amthor, Daniel Fischer, Winfried E and Dirk Stelzer [11]. This research proposes a conceptual design for threat reaction methodologies, security architectures and processes, data representation requirements, and the first steps toward integrating threat intelligence sharing platforms with security-policy-controlled systems [11]. Overall, the above articles match the search criteria which Author is looking for.

This research investigates the role of information Technology in cyberwar. The research question is as follows:
RQ: In what ways does information technology play a role in cyberwarfare?

## 3 METHODOLOGY

In this exploratory study. The Author perfoemed systematic literature review as presented in the guidelines by Amjad Abu Hassan, Mohammad Alshayeb and Lahouari Ghouti and was implemented [12]. A  systematic Literature Review is an report that distinguishes, assesses, unites, and gathers the primary studies on specific research issues [15]. By doing a ressearch in light of past important examinations, a precise systematic Literature Review turns into a helpful way of acquiring a solution. The objective of a systematic Literature Reviewis, is, to sum up, earlier exploration, decide the flaws that should be filled among past and flow research, compose a report, and make an examination structure.3.1 Study Strategy:

A comprehensive process to expand coverage, automatic and human searches were used to find peer-reviewed articles that have been published.. The search criteria used two databases ACM Digital library [14] & IEEE Explore [13]. The ACM and IEEE Explore are the two the most extensive datasets of research in information technology [16], and hence there were selected to conduct the study.

The author used derived words to search various search strings to obtain as many relevant primary research publications as possible. The Author used Boolean operations such as 'AND,' 'OR,' etc., to develop a search strategy and select the articles that match the research keywords. To test the effectiveness of locating good papers using each string, the Author prepared a few different search queries utilizing terms obtained from the synthesized Research Question (RQ). The search string was primarily based on critical time present in my formulated Research Question and research topic. The Search terms used were ("cyberwar" or "cyber warfare" or "Cyber security war" or " Malware" or " Cyber Attack") AND ("Information Technology" or "Technology-driven"). A few tweaks were made to meet the syntax of the search engines utilized in this study as needed.

### IEEE: (("cyberwar" or "cyber warfare" or "Cyber security war" or " Malware" or " Cyber Attack") AND ("Information Technology" or "Technology-driven"))

ACM: ("cyberwar" or "cyber warfare" or "Cyber security war" or " Malware" or " Cyber Attack") AND ("Information Technology" or "Technology-driven")

3.2 Study Selection

3.2.1 Inclusion Criteria:

This study includes papers about cyberwarfare attacks using information technology authored and published in a string search from January 1, 2000, to April 1, 2022, and this review contains some research findings that help in conducting research work. The related articles were screened for relevance in the final selection process.

- Are the relevant keywords found in the title for the need of the study
- Are the papers written in English
- Are the papers peer-reviewed
- Are the research paper falls in the publication year of 2000 and 2022

3.2.2 Exclusion Criteria:

- If cyber warfare was not a primary focus of the selected papers.
- Masters and Ph.D. research that haven't been published in peer-reviewed journals or conferences.
- Not formatted articles

3.2.3 Quality Assessment:

Despite the fact that all of the papers were sourced from reputable databases, the author want to ensure that the studies chosen for this study are relevant to the topic. An assessment criterion, based on Barbara Kitchenham's guidelines [15], is used to review the quality of the papers. The structure of the articles is introduction, methodology, results, and discussion; hence it is easy to decide which one should be considered for the primary study. Table 1 shows the quality criteria used to review the quality of the retrieved research papers, and the author has used a "Yes" or "No"  measure to assess the quality of the articles.

Table 1: Quality criteria of the selected papers

| S.NO | Quality Criteria | Article-1 | Article -2 | Article -3 | Article -4 | Article -5 |
|---|---|---|---|---|---|---|
| 1 | Introduction matches the keywords | YES | YES | YES | YES | YES |
| 2 | The methodology section is referred to correctly in the articles | YES | YES | YES | NO | YES |
| 3 | Papers discussed the research question of our study | YES | NO | YES | YES | YES |
| 4 | Reports concluded with the desired outcome | NO | YES | NO | YES | YES |

3.3 Study Procedure

On April 3, 2022, the search for research publications began, with the electronic databases ACM Digital Library and IEEE Xplore being used as recommended. Then, utilizing keywords like Cyberwarfare, cyberwar, and information technology, inclusion and exclusion criteria such as papers published more minor than the year 2000 are applied to the title of the research item in the search conditions. Table 2 shows no articles in the ACM and IEEE related to the search criteria related to the research question. We utilized simple random sampling to select the list of articles since we kept all the pieces in a spreadsheet, assigned random numeric values, sorted most significant to most minor, and later established the top five papers from the spreadsheet.

Table 2: No Articles related to search criteria

| S.No | Search criteria | No of the Articles in IEEE | No of the Articles in ACM |
|---|---|---|---|
| 1 | Cyberwars or Cyberattacks | 177 | 218 |
| 2 | Information Technology and Cyber War | 53 | 56 |

Selective research was entered into a data extraction sheet using a Microsoft Spreadsheet. Details such as the year, the algorithm employed, types of cyber wars, and data sources were discovered during the investigations. This study uses a qualitative approach and conducts a thematic analysis of the five research papers included in the study list to identify important topics that illustrate a variety of cyber warfare. Most of the subjects were discovered by thematic analysis of abstracts; however, a few themes were uncovered by reading the introduction and results portions of research papers to supplement the findings.

3.4 Data Extraction

This study used thematic analysis to extract data from the Study List prepared by analyzing the text of the abstract of each article. Based on the study conditions and after applying simple random sampling to the spreadsheet, the Author gathered specific research papers and now will start to examine this content to see if it fulfills the research study. After applying the Inclusive and Exclusive criteria, duplicates are deleted.

The Author gathered all relevant studies before writing the research paper; however, this does not imply that we can identify the final list of documents because the quality of all selected studies is challenging to be desirable. To solve this, the Author used a quality criterion technique and meticulously evaluated each paper's quality by reading the entire abstract and numerous other standardized publications for definitions.

**4. RESULTS**

The articles were gathered in a spreadsheet and formatted to start a thematic analysis process to get the desired results for the research study. The author checked the reports for any irrelevant data or misleading data, making the research study off track. Such type of data can influence the results in the wrong way. The Author concentrated on information from journals and reference papers relating to information technology and cyber warfare, and the publication outlet, which was the primary source of these findings, was recorded. First, the Author subjectively presented the data, including the abstract, keywords, research studies, and so on, to characterize the features of the selected research studies. The Author compiled all the research papers into a single sheet and used thematic analysis to divide the data into themes and definitions, followed by the frequency of occurrences associated with the pieces using a qualitative analysis approach.

A group of fifteen raw data values was formed, and those were studied by assigning each theme a definition and a frequency showing the percentage of each piece in the entire set of fifteen raw data values collected. After analyzing the abstract of the articles, around five themes have been identified, giving each theme a definition and calculating the frequency of how many pieces fall under the specific them mentioned in table 3. Now, using the extracted data and all the numbers from the journals, we could locate and generate distinct frequencies that supported our study topic.

Table 3: Thematic Analysis of the Raw Data

| Themes | Description | Frequency |
|---|---|---|
| Economic Disruption | Attacking the economy by cyber attacks | 2 |
| Data Breach | A data breach occurs when sensitive or confidential data is copied without permission. | 5 |
| Malware | Malware is a blanket word for any malicious software designed to harm or exploit any programmable device. | 2 |
| Social, Behavioral Aspects | Articles related to Social, Behavioral Aspects | 1 |
| Legal Aspects | Attacking the legal aspects of an organization | 1 |
| Cyber Attacks | It is an attempt to abuse or exploit another person's computer physically. Hackers are attempting to gain access to consumers' personal information. | 3 |
| Vulnerability | Being exposed to the danger of being attacked or damaged is a trait or state. | 1 |

We did perform the thematic analysis, and we did get various themes such as Economic Disruption, Data breach, Legal Aspects and Cyber Attacks. Between the years 2000 and 2022, information technology played a critical part in the rise of cyber-attacks around the world, according to this study. We did research the themes and frequencies in the graph below. The thematic analysis suggests that the number of articles published was related to the data breach theme in the digital databases.

Graph 1: Thematic Analysis of the Raw Data

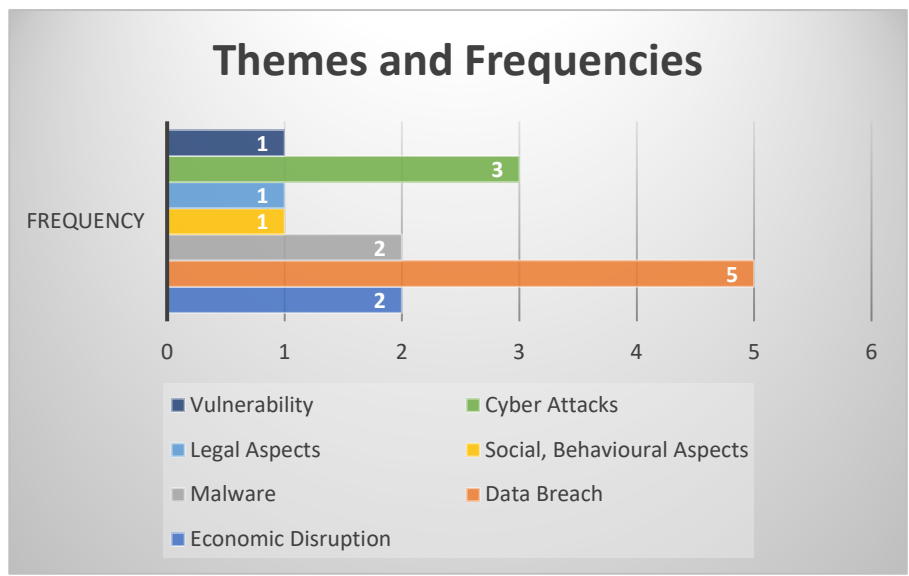

## 5. DISCUSSION

The Author evaluates the project's research topic findings of how information technology is employed in cyberwar, gives the results, and explains the Systematic literature review's imitations. The author recommended an additional investigation, and accurate testing revealed that the information technology platform contains various data breaches, worms, trojans, and phishing threats that could lead to attacks, jeopardizing the platform's integrity.

The study's limitations fully detail the empirical research of various search algorithms in various electronic databases, such as IEEE Xplore and the ACM digital library, while neglecting other unnecessary database components. As a result, certain search conditions were skipped over. After careful research, the author discovered that many of the available frameworks lack methodology and data analysis, resulting in less accurate outcomes even though several studies show a wide range of metric performance.

Cyber Warfare and Cyberattacks, Malware, Cyber Threats, and Data Breach, were four common occurrences. According to a sample of five publications, these strategies are the most regularly used, followed by other frequencies. As a result of this finding, we may deduce that data breaches and cyber-attacks are the most common occurrences in the research study. However, using only five papers as a sample has disadvantages because it is a limited sample, and many papers linked to cyber warfare with more helpful technologies may be missed in the random sampling process and appropriate content from better journals. The qualitative analytical approach has several drawbacks that can be justified by credibility, confirmability, transferability and dependability.

To check that the results were valid, the authors evaluated the raw data taken from the results to the final set of frequencies and concluded that both values were equal. Study strategy, study inclusion, study selection, exclusion, quality evaluation, study technique, and data extraction were all followed according to the systematic literature review methodology. And Author examined if the extracted data is relevant or not, whether it is incomplete, and we can claim the data findings are internally authentic after taking all of this into account and performing a significant quality evaluation. The proximity similarity model was used for external validation, and the findings with the two dominant frequencies were applied to the rest of the population that was omitted. The outcomes might or might not be equivalent. As stated in the preceding paragraphs, we selected only five research as a sample, which cannot be confirmed when the results are published in new journal publications because the predominant frequencies are connected to Information technology.